\documentclass[aps,preprint, showpacs]{revtex4}
\input{psfig.sty}

\usepackage{amsmath,bm,graphics}
\newcommand{\csch}{\textrm{ csch }}

\newcommand{\arccot}{\textrm{ arccot }}
\newcommand{\arccoth}{\textrm{ arccoth }}
\newcommand{\e}{\textrm { e}}

\newcommand{\be}{\begin{eqnarray}}
\newcommand{\ee}{\end{eqnarray}}

\begin{document}

\title{The Trigonometric 
Rosen-Morse Potential in the Supersymmetric Quantum Mechanics
and its  Exact Solutions}

\author{C.\ B.\ Compean and M.\ Kirchbach}

\affiliation{Instituto de F\'{\i}sica, \\
          Universidad Aut\'onoma de San Luis Potos\'{\i},\\
         Av. Manuel Nava 6, San Luis Potos\'{\i}, S.L.P. 78290, M\'exico}

\date{\today}
\begin{abstract}
The analytic solutions of the one-dimensional Schr\"odinger equation 
for the trigonometric Rosen-Morse potential reported in the literature
rely upon the Jacobi polynomials with complex indices  and
complex arguments. We first draw attention to the fact that the
complex Jacobi polynomials have non-trivial orthogonality properties
which make them uncomfortable for physics applications.
Instead we here  solve above equation  in terms of real orthogonal
polynomials. The new solutions are used in the construction of the 
quantum-mechanical superpotential. 
\end{abstract}

\maketitle

\section{Introduction}
Supersymmetric quantum mechanics was originally proposed by 
Witten \cite{Witten}
as a simple learning ground for the basic concepts of supersymmetric quantum 
field theories but soon after it evolved to a research field on its own rights.
Supersymmetric quantum mechanics starts with the factorization of
one-dimensional Hamiltonians, 
\begin{eqnarray}
H(y) &=&-\frac{\hbar^{2}}{2m}\frac{d^{2}}{dy^{2}}+V(y),
\label{1dim_H}
\end{eqnarray} 
according to $H(y)=A^+(y) A^-(y)+ \epsilon $  with $A^\pm(y) =
\left( \pm\frac{\hbar}{\sqrt{2m}} \frac{d}{dy} +
U(y)\right)$ where $U(y)$ is the superpotential.
Next it proves that if $\psi_n(y)$ is an exact solution of the
$H(y)$ eigenvalue problem, $H(y) \psi_{n} (y)=E_n\, \psi_{n}(y)$,  
then $A(y)\psi_n(y)$ is an eigenfunction to 
$\widetilde{H}(y)=A^-(y)A^+(y) +\epsilon  $ 
corresponding to same eigenvalue. In other words,
knowing the superpotential allows to generate
the $\widetilde{H}(y)$ spectrum from the spectrum of 
$H(y)$ and vise versa. 
Moreover, in case of zero ground state (gst) energy,
knowing $U(y)$ allows to solve $A^-(y)\psi_{{\mbox{gst}}}(y)=0$
and obtain the ground state wave function. In other words,
knowing the ground state wave function allows to recover
the superpotential as 
\begin{equation}
U(y)=-\frac{\hbar }{\sqrt{2m}}\frac{d}{dy}\ln \psi_{\mbox{gst}}(y)\, .
\label{super_pt}
\end{equation}
At that stage one uses the isospectral pair $H(y)$, and $\widetilde{H}(y)$
in the construction of  the super-Hamiltonian {\bf H}(y) as
{\bf H}(y)=diag$\left( H(y),\widetilde{H}(y)\right)$ and upon introducing 
 ``charges'' as
\begin{equation}
Q(y)=\left(
\begin{array}{cc}
0&0\\
A^-(y)&0
\end{array}
\right)\, , \quad
Q^\dagger (y) =\left(
\begin{array}{cc}
0&A^+ (y)\\
0&0
\end{array}
\right)\, ,
\label{susy_charges}
\end{equation}
proves them to satisfy the following algebra:
\begin{eqnarray}
\lbrace Q (y),Q^\dagger(y) \rbrace ={\mathbf H}(y),
&\quad& \lbrace Q(y),Q(y) \rbrace =
\lbrace Q^\dagger (y) ,Q^\dagger (y) \rbrace=0,
\nonumber\\
\lbrack Q(y),{\mathbf H}(y)\rbrack  &=&
\lbrack Q^\dagger(y), {\mathbf H}(y)\rbrack =0\, .   
\end{eqnarray}
The relationship to the field-theoretic SUSY is then established 
through the observation that in case $A^-(y)\psi_{\mbox{gst}}(y)\not=0$
then the charges do not annihilate the respecive vacua,
$Q(y)\psi_{\mbox{gst}}(y)\not=0$, and $Q^\dagger(y)
\widetilde{\psi}_{\mbox{gst}}(y)\not=0$, and  SUSY is spontaneously broken.
On the contrary, when  $A^-(y)\psi_{\mbox{gst}}(y)=0$,  then the charges 
annihilate the ground states  which 
is equivalent to the absence of charge condensates there,
and thereby to SUSY realization
in the manifest (multiplet) Wigner-Weyl mode.  
In this manner supersymmetric quantum mechanics relates to 
SUSY in field theory where the r\'ole of $Q(y)$ and $Q^\dagger(y)$ is
taken by boson-fermion (and vise versa) ladder operators.

\noindent
Above considerations clearly reveal importance of knowing the exact 
solutions of the quantum mechanics Hamiltonians. These solutions are 
furthermore important in the construction of higher dimensional 
charge algebras with more but two Hamiltonians (hierarchy of Hamiltonians)
\cite{Sukumar}.
The supersymmetric quantum mechanics manages  a family of exactly soluble
potentials \cite{Khare}, one of them being the trigonometric 
Rosen-Morse potential (tRMP).
As long as this potential is obtained from the Eckart potential 
\cite{Eckart} by complexification of the argument and one of
the constants, also  its solutions have been concluded  from
those of the Eckart potential by same procedure.
In so doing, one ends up with Jacobi polynomials with complex
indices  and complex arguments. However, such complex polynomials
are not comfortable for physical applications basically 
in view of their non-trivial orthogonality properties 
\cite{Jacobi-c,Jacobi-otros}.

\begin{quote}
We here make the case that the trigonometric Rosen-Morse potential
is exactly soluble in terms of a family of {\it real }
orthogonal polynomials and present the solutions. 
\end{quote}

The paper is organized as follows. In the next Section we present the 
tRMP derivation form the Eckart potential and draw attention to
the non-trivial orthogonality properties of the
Jacobi polynomials with complex parameters and arguments. In Section III
we solve analytically the one-dimensional Schr\"odinger equation with the
trigonometric Rosen-Morse potential and present the solutions.
In Section IV we employ the exact ground state wave function in the
construction of the tRMP superpotential.
The paper closes with  brief concluding remarks.

\section{The trigonometric Rosen-Morse potential as complexified
Eckart potential. }

\begin{figure}[htbp]
\centerline{\psfig{figure=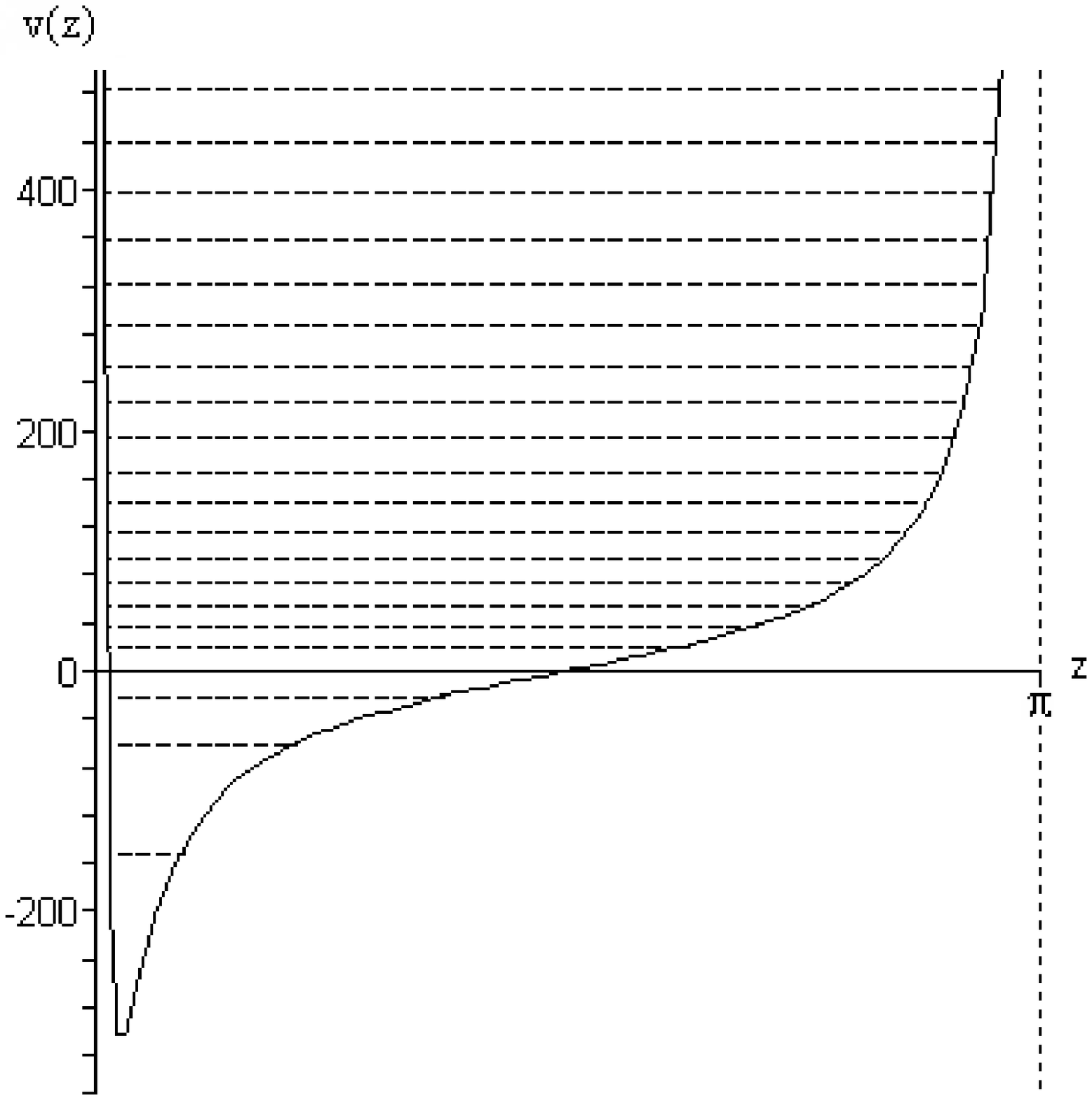,width=7cm}}
\vspace{0.1cm}
{\small Fig.~II.
\hspace{0.2cm} 
The trigonometric Rosen-Morse potential.
The solid line represents the potential, while the dashed lines
are the energy levels. Notice that the argument of this
potential is bound between $0< z <\pi $ 
and the number of states is unlimited. The potential parameters take 
the values $a=1$, $b=50$.
}
\label{fig-RMt}
\end{figure}

Before proceeding further we first introduce a properly chosen
length scale $d$ and  change variables  in the
one--dimensional Schr\"odinger equation (\ref{1dim_H}) to dimensionless 
ones according to
\begin{equation}
z=\frac{y}{d},\quad v(z)=V(dz)/(\hbar^{2}/2md^{2}),
\quad \epsilon_n =E_n/ (\hbar^{2}/2md^{2})\,.
\end{equation}
Next we  employ the  Eckart potential \cite{Eckart},
\be v(z)=-2 b \coth z +a(a-1)\csch^2 z\ ,
\label{v-Eckart}\ee where $b>a^2 \,$ .
The exact solutions to the Eckart potential read:
\begin{eqnarray} 
\psi_n(x)&=&c_n (x-1)^{(\beta_n-n-a)/2} (x+1) ^{-(\beta_n+n+a)/2}\,  
P_n^{ \beta_n-n-a, -(\beta_n+ n+a) }(x) \,, \nonumber\\
\quad x=\coth z\, , & \quad &\beta_n=\frac{b}{n+a}\, .
\label{psi-Eckart1}\end{eqnarray}
Here, $P_n^{(\beta_n-n-a,-(\beta_n+n+a))}(x)$ are the
well known Jacobi polynomials \cite{Dennery},\cite{handbook} 
with  $n\leq (b^{1/2}-a)$, and  $c_n$  is a normalization constant.  
Equation 
(\ref{psi-Eckart1}) equivalently rewrites to
\be \psi_n(x)=c_n (x^2-1)^{-(n+a)/2} \e^{-\beta_n\ \arccoth x}
 P_n^{(\beta_n-n-a,
-(\beta_n+n+a))}(x) \ .\label{psi-Eckart2}\ee
The corresponding energy value spectrum is determined by  
\be \epsilon_n=-(n+a)^2-{b^2\over (n+a)^2}\ .\label{ep-Eckart}\ee
{}Form and energy spectrum of the Eckart potential are illustrated by 
Fig.~ I.
\begin{figure}[htbp]
\centerline{\psfig{figure=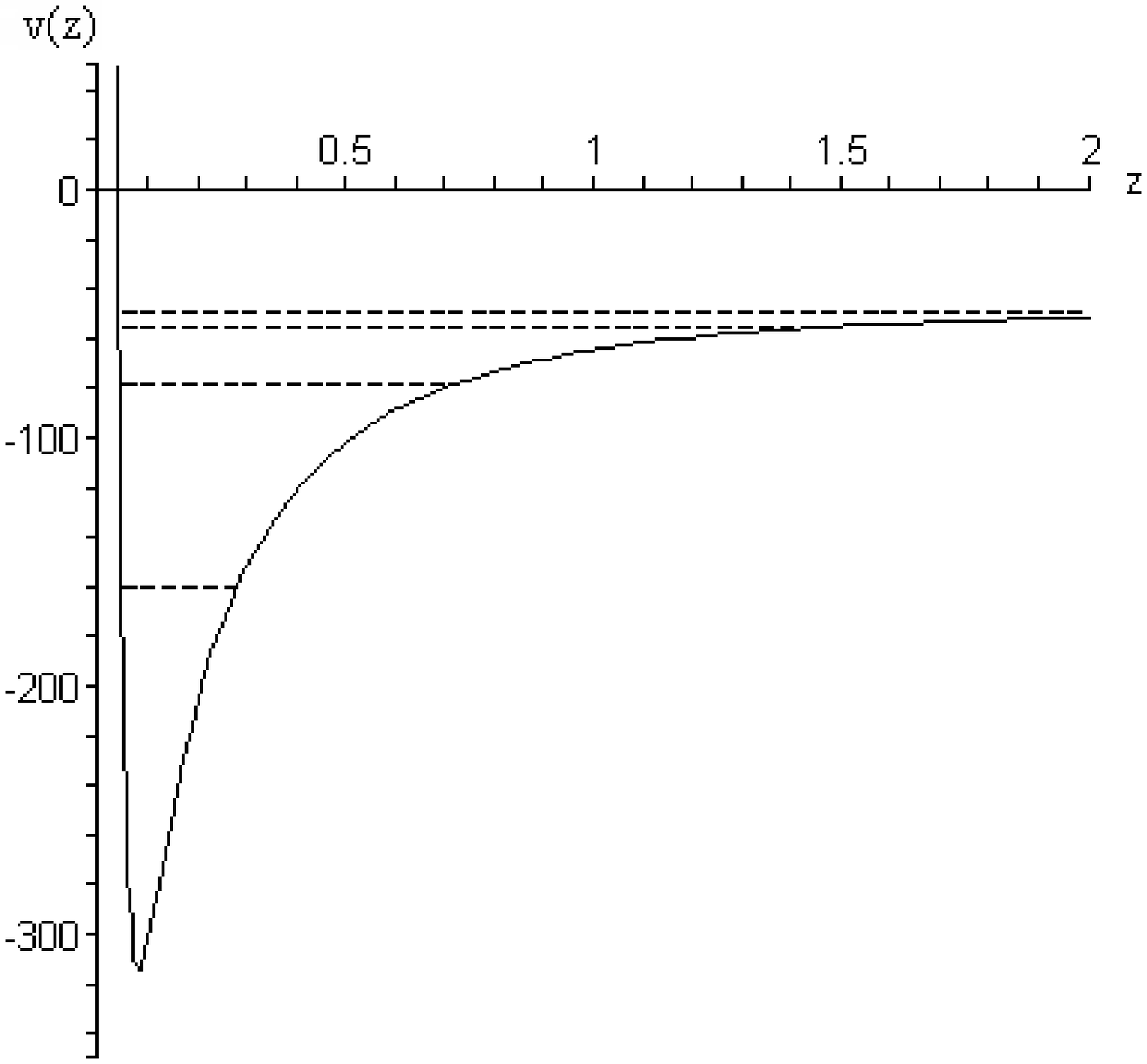,width=7cm}}
\vspace{0.1cm}
{\small Fig.~I. \hspace{0.cm} 
Eckart potential. The solid curve represents the
potential while the dashed lines are the
energy levels. Notice that the argument of this potential is 
unbound from above, i.e. $0<   z < \infty $ and the
number of bound states limited. The parameters of the displayed potential
take  the values $a=-1$, $b=50$.} 
\label{fig-Eckart}
\end{figure}
Let's now complexify the argument of the Eckart potential and
one of its constants according to
\be z \rightarrow -i\ z\ ,\quad \mbox{or, equivalently,} \quad 
x\longrightarrow ix\, ; \quad b\to ib.
 \label{z-iz}\ee
Substitution of Eq.~(\ref{z-iz}) into (\ref{v-Eckart}) results in
\be v(z)=-2 b \cot z +a(a-1)\csc^2 z\, , \label{v-RMt}\ee
and thereby in the trigonometric Rosen-Morse potential \cite{Khare}
shown on Fig.~\ref{fig-RMt}.\\
In the literature \cite{Khare} the solution of the Schr\"odinger equation
with the trigonometric Rosen-Morse potential is concluded from 
Eq.~(\ref{psi-Eckart2}) through complexification of $b$ and $x$ leading to
\be \psi_n(ix)=c_n ((ix)^2-1)^{-(n+a)/2} \e^{-\beta_n\arccoth\,  ix } 
P_n^{((i\ \beta_n-n-a),
-(i\beta_n+n+a))}(ix) \, .\label{psi-RMt}\ee
In other words, these solutions need the Jacobi polynomials with
complex indices and complex arguments.
Unfortunately, the complex  Jacobi polynomials are  not very  well suited
for physics applications. Suffices to write down these polynomials 
explicitly,
\begin{eqnarray}
P_n^{(A,B)}(ix)
= \frac{2^{n}(A+1)_{n}}{(n+A+B+1)_{n}} && 
_{2}F_{1}\left( -n, n+A+B+1; A+1\vert \frac{1-ix}{2}\right)\,,
\nonumber\\
_{2}F_{1}\left( -n, n+A+B+1;A+1\vert \frac{1-ix}{2}\right)
&=&
\frac{\Gamma (A+1)}
{\Gamma (n+A+B+1)\Gamma (-(n+B))}\nonumber\\
& & \int_{0}^{1}
t^{n+A+B}(1-t)^{-(n+B+1)}
\left(1-t\frac{1-ix}{2}\right )^{n}dt\, ,\nonumber\\
A=i\beta_n-n-a\, ,&\quad& B=-(i\beta_n +n+a)\, ,
\label{Jacobi_cmplx}
\end{eqnarray}
where $_{2}F_{1}(a,b;c|x)$ is the well known hypergeometric function,
and $(...)_{n}$ is  the Pochhammer symbol of the expression in
the parentheses, in order to become aware of the calculational
difficulties to be expected. It is obvious that one has to worry 
about the interplay between indices and integration contours,  
a subject studied in  \cite{Jacobi-c,Jacobi-otros}.
There the authors claim dependence of the orthogonality properties on 
the indices and  the integration contours. 
In order to avoid all those difficulties we here 
search for real solutions of the one-dimensional Schr\"odinger 
equation with the trigonometric Rosen-Morse potential.

\section{The one-dimensional Schr\"odinger equation for the
trigonometric Rosen-Morse potential. \label{Secc-s-Sch}}

\subsection{\label{Secc-Rodrigues}The Sturm-Liouville method and the 
Rodrigues formula.}

{}For the sake of self-sufficiency of the representation we here review in
brief the basics of the Sturm-Liuoville technique for solving
second order differential equations.

The method of Sturm-Liouville \cite{wikipedia} applies to differential
equations of the type 
\be {d\over d\ x}\left(p(x) {d\ y\over d\ x}\right)+q(x)\ y=\lambda w(x)\ 
y\ ,\label{S-L}\ee
and searches for the $\lambda$ values that allow for a solution. 
The solution of Eq.~(\ref{S-L}), where  $w(x)$ stands for the ``weight'', 
or, ``density'' function, are eigenfunctions of a
Hermitian differential operator on the space of functions
defined by the initially conditions.
Some special cases of the Sturm-Liouville type of differential equations
 allow for solutions by means of the  Rodrigues formula \cite{Dennery}. 
To be specific, for  $q(x)=0$, and  $p(x)=w(x)s(x)$, 
where  $s(x)$ at most a second order polynomial,
the solution of Eq.~(\ref{S-L}) is given by a family of orthogonal
polynomials. The classical polynomials of Hermite,
Laguerre, Legendre and Jacobi are prominent examples for that.\\

In order to create the  orthogonal polynomial solutions
one considers the function \cite{Dennery}
\be C_m(x)=\frac{1}{w(x)}{d^m\over d\ x^m}(w(x)\ s(x)^m) \ ,
\label{Rodrigues-0}\ee
with $C_1(x)$, $w(x)$, and $s(x)$ 
satisfying the following conditions:
\begin{enumerate}
\item $C_1(x)$ is a polynomial of first order,
\item $s(x)$ is a polynomial of at most second order and  real
roots,
\label{cond2-Rodrigues}
\item $w(x)$ is real, positive and  integrable within a given interval
 $[a,b]$, and satisfies the boundary conditions
\be w(a)s(a)= w(b)s(b)=0 \ .
\label{rule_tobreak}\ee
\end{enumerate}
Above three conditions seem quite restrictive indeed but nonetheless
they allow for the construction of all the {\it classical \/}
orthogonal polynomials reported in the standard textbooks 
\cite{Dennery}, \cite{handbook},
\cite{wikipedia}. 
The table  shows all the ingredients
of the Rodrigues formula required for the construction of the respective
orthogonal polynomials.\\

\begin{table}[htbp]
\caption{ Special functions and their characteristics.}
\vspace*{0.21truein}
\begin{tabular}{||l|c|c|c|c|l||}
\hline \hline
Name & Symbol & $w(x)$ & $s(x)$ & Interval & Conditions \\
\hline \hline
Hermite & $H_m(x)$ & $\e^{-x^2}$ &1 & $(-\infty,\infty)$ &\\
\hline
Laguerre & $L_m^\nu(x)$ & $x^\nu\e^{-x}$ & $x$ & $[0,\infty)$ & $(\nu>-1)$\\
\hline
Jacobi & $P_m^{(\nu, \mu)}(x)$ & $(1-x)^{\nu}(1+x)^{\mu}$ & $(1-x^2)$ & 
$[-1,1]$ & $(\nu,\mu>-1)$\\
\hline
Gegenbauer & $C_m^\lambda(x)$ & $(1-x^2)^{\lambda-1/2}$ & $(1-x^2)$ & 
$[-1,1]$ & $(\lambda>-1/2)$\\
\hline
Legendre & $P_m(x)$ & $1$ & $(1-x^2)$ & $[-1,1]$ & \\
\hline
Chebyshev, Type I & $T_m(x)$ & $(1-x^2)^{-1/2}$ & $(1-x^2)$ & $[-1,1]$ & \\
\hline
Chebyshev, Type II& $U_m(x)$ & $(1-x^2)^{1/2}$ & $(1-x^2)$ & $[-1,1]$ & \\
\hline \hline
This work  & $C_{m+1}^{(a,b)}(x)$ & $(1+x^2)^{-\mu}\e^{-2{b\over \mu}
\cot^{-1} x}$ & $(1+x^2)$ & $(-\infty,\infty)$ & $(\mu=m+1+a)$\\
\hline \hline
\end{tabular}
\end{table}
If in addition one demands an orthonormalized set of polynomials,
one has to introduce into Eq.~(\ref{Rodrigues-0}) 
an additional constant, here denoted by $K_{m}$, according to
\be C_m(x)={1\over K_m\ w (x)}{d^m\over d\ x^m}\left(w(x)\ s(x)^m\right) \ .
\label{Rodrigues-1} \ee
In terms of $C_{m}(x)$, the Sturm-Liouville equation takes the form
\be
{d\over d\ x}\left(w(x)\ s(x) 
{d\ C_m(x)\over d\ x}\right)=-\lambda_m w(x)\ C_m(x)\ ,
\label{d2-R1}\ee
or, equivalently,
\be s (x){{d^{\, 2}C_m(x)} \over {d\ x^2}}+
{1\over {w(x)}}\left({{d\ s(x)w(x)}\over {d\ x}}\right){d\ C_m(x)
\over d\ x}+\lambda_m \ C_m(x)=0\ ,\label{d2-R2}
\ee
where 
\be \lambda_m=-m\left(K_1{{d\ C_1(x)} \over {d\ x}}+
{1\over 2}(m-1) {{d^{\, 2} s(x)}\over 
{d\ x^2}}\right)\ .\label{lamb}\ee
Notice that
\be C_1 (x) = {1\over {K_1w(x)}}
\left({{d\ s(x)w(x)}\over {d\ x}}\right)\ .\label{C1-R}\ee
All classical polynomials can be obtained through above procedure and 
vise versa. The polynomials that can be obtained from the Rodrigues
formula and which satisfy the three conditions mentioned above
are necessarily the classical orthogonal polynomials, a result
due to Ref.~\cite{Cryer}.

\subsection{Solving Schr\"odinger's equation for
the trigonometric Rosen-Morse potential.}

In this Section we present the solution of the one-dimensional
Schr\"odinger equation for the trigonometric Rosen-Morse potential
as obtained in Ref.~\cite{Cliff_tesis} and 
without any reliance on the complex Jacobi polynomials.
{}For this purpose we first have to reshape Schr\"odinger's equation
\be {d^{\, 2}\ R(z) \over d\ z^2}+\left(
2 b\cot z-a(a+1)\csc^2 z+\epsilon\right) R(z)=0\, ,
\label{Sch-RMt}\ee
to the Sturm-Liouville form in Eq.~(\ref{d2-R2}).
To do so we first change variables to
\be R(z) =\e^{-\alpha z/2}F(z) \ ,\ee
with $\alpha$ being a constant  and then substitute  
in Eq.~(\ref{Sch-RMt}). After some simple algebraic
 manipulations one arrives at
\be {d^{\, 2}\ F(z)\over d\ z^2}-{\alpha}{d\ F(z)\over d\ z}+
\left(2 b\cot z-a(a+1)
\csc^2 z + \left(\left({\alpha\over 2}\right)^2+\epsilon\right)\right)F(z) 
= 0\ .
\label{Sch-RMt2}\ee
Changing once more variables to  $x=\cot z$ in which case
$F(z)$ becomes a function of $x$ denoted by $f(x)$, i.e. 
\be  F(z)\rightarrow f(x)\ ,\quad  x=\cot z\, ,\ee
Eq.~(\ref{Sch-RMt2}) takes the form
\be 
\left(2 bx-a(a+1)(1+x^2) + \left(\left({\alpha\over 2}\right)^2+\epsilon
\right)\right)f (x) 
&+& (1+x^2)^2{d^{\, 2}\ f(x)\over d\ x^2}\nonumber\\
&+&2(1+x^2)
\left({\alpha\over 2}+x\right) 
{d\ f(x) \over d\ x} = 0\ \, .
 \label{Sch-RMt3}\ee
Finally,  upon substituting the factorization ansatz
\be f(x)=(1+x^2)^{-(1-\beta)/2}C(x) \ ,\ee
into Eq.~(\ref{Sch-RMt3}) and a subsequent division by $(1+x^2)^{(1+
\beta)/2}$ one finds as intermediate result the equation
\be \left((-\beta(1-\beta)-a(a+1)) +{(-\alpha(1-\beta)+2 b)x + \left(
\left({\alpha\over 2}\right)^2-(1-b)^2+\epsilon\right)\over (1+x^2)}\right)
C(x)  \nonumber \\
+ (1+x^2){d^{\, 2}\ C(x)\over d\ x^2}+2\left({\alpha\over 2}+\beta x\right) 
{d\ C (x) \over d\ x} & = & 0\ .\nonumber\\
\label{Sch-RMt4}\ee
If  Eq.~(\ref{Sch-RMt4}) is to coincide in form with Eq.~(\ref{d2-R2}), 
following conditions have to be fulfilled:
\be
-\alpha(1-\beta)+2 b=0\ ,\label{ab-b-1}\\
\left({\alpha\over 2}\right)^2-(1-\beta)^2+\epsilon=0\label{ep-b2_a2-1}\ .
\ee
Substitution of Eqs.~(\ref{ab-b-1}), (\ref{ep-b2_a2-1}) into
Eq.~(\ref{Sch-RMt4}) amounts to
\be (1+x^2){d^{\, 2}\ C(x) \over d\ x^2}+2\left({\alpha\over 2}+\beta x\right) 
{d\ C(x)
\over d\ x} + (-\beta(1-\beta)-a(a+1))C(x) = 0\ . \label{Sch-RMt5}\ee
Equation~(\ref{Sch-RMt5}) relates to Eqs.~(\ref{d2-R2})--(\ref{C1-R}) 
via
\be -\beta(1-\beta)-a(a+1) = -m(2\beta+m-1)\ .\label{b-1}\ee
In now determining $\beta $ from Eq.~(\ref{b-1}), substituting it into 
Eqs.~(\ref{ab-b-1}), (\ref{ep-b2_a2-1}), and  
shifting $m$ to   $m\rightarrow n-1$, the following $n$ dependent
constants are found:
\be
\beta_n=-(n+a)+1\ ,&\quad& \alpha_n={2 b\over n+a}\ ,\\
\epsilon_n &=& (n+a)^2-{b^2\over (n+a)^2}\ ,
\ee
with  $n\ge 1$.\\
In this way one encounters $w(x)$ and $s(x)$ as
\be 
w_n(x)&=&(1+x^2)^{-(n+a)}\e^{-{\alpha_n} \arccot x}\ ,\\
s(x)&=&1+x^2\ .
\ee
The polynomials which resolve the tRMP Schr\"odinger equation are now 
obtained in exploiting the Rodrigues formula (\ref{Rodrigues-1}) 
when rewritten in terms of  $n$  as
\be C^{(a,b)}_{n}(x)={1\over K_n\ w(x)}{d^{n-1}\over d\ x^{n-1}}\left(w_n(x)\ 
s(x)^{n-1}\right) \, . \label{pol-nvo}\ee
The lowest $C_n^{(a,b)}(x)$ polynomials obtained in this fashion  read
\be
C^{(a,b)}_{1}(x)&=&{1\over K_1}\ , \\
C^{(a,b)}_{2}(x)&=&{2\over K_2}\left(-(1+a)x + {b\over 2+a}\right)\ , \\
C^{(a,b)}_{3}(x)&=&{2\over K_3}\left((1+a)(2a+3)x^2-2(2a+3){b\over 3+a}x+ 
\left({2b^2\over (3+a)^2}-(1+a)\right)\right)\ , \\
C^{(a,b)}_{4}(x)&=&{4\over K_4}
\left(-(1+a)(2a+3)(2+a)x^3 + 3(a+2)(2a+3){b\over (4+a)} x^2\right. \nonumber 
\\
& &\left.  -3(2+a)\left(2 {b^2\over (4+a)^2}-(1+a) \right)x + \left({2b^3\over 
(4+a)^3}-(3a+4){b \over 4+a}\right) \right) \ ,\\
C^{(a,b)}_{5}(x)&=&{4\over K_5}\left((1+a)(2a+3)(2+a)(2a+5)x^4 - 4(2a+3) (2+a) 
(2a+5) {b\over (5+a)} x^3 \right.  \nonumber \\
& &\left. +6(2+a) (2a+5) \left({2 b^2\over (5+a)^2}-(1+a) \right)x^2 \right.  
\nonumber \\
& & - 4 (2a+5) \left({2b^3\over (5+a)^3}-(3a+4){b \over 5+a}\right)x \nonumber 
\\
& &\left. +\left({4b^4\over (5+a)^4}-{4b^2\over (5+a)^2}(3a+5)+3(2+a)(1+a)
\right) \right) \ ,
\label{C-pol}
\ee
where $x=\cot z$.\\
Above polynomials solve exactly Eq.~(\ref{Sch-RMt5}) which can be immediately
cross-checked  by back-substituting Eqs.~(\ref{C-pol}) 
into Eq.~(\ref{Sch-RMt5}). Employing symbolic mathematical programs is 
quite useful in that regard.
\begin{quote}
Notice that the solution was found under less rigid requirements but
the ones listed immediately after Eq.~(\ref{Rodrigues-0}) above. Indeed,
\begin{itemize}
\item the roots of our $s(x)$ function are imaginary (this is the only place
where the complexification of the Eckart potential seems to have left
footprints),
 \item  Equation~(\ref{rule_tobreak}) turned out to be
a rule that allows to be broken.
\end{itemize}
In that sense, the orthogonal functions found here must belong to a different
class of orthogonal functions.
\end{quote}

Their orthogonality  is obtained as
\be \int_{-\infty}^\infty {dx\over s(x)}(w_n(x))^{1/2}C_n^{(a,b)} 
(w_{n'}(x))^{1/2} C_{n'}^{(a,b)}=\delta_{n\ n'}\ .\label{orto-1}\ee
Equation  (\ref{orto-1}) shows convincingly that the new solutions
have well defined orthogonality properties on the real axes, which
qualifies them as comfortable wave functions in quantum mechanics
applications. 

The orthogonality condition for the wave functions $R_{n}(z)$ reads
\be \int_0^\pi dz\ R_{n}(z) (R_{n'}(z))^{*} =\delta_{n\ n'}\, .\ee
A version interesting for physical application (see concluding Section) 
is the one with $a=0$, 
\be
v(z)=-2b\cot \, z\, ,
\label{RVM_pot}
\ee
in which case the normalization constant is calculated as 
\be K_n=\left({(n!)^2 n^3(1-\e^{-2\pi b/n})\over 4 b (b^2+n^4)}\right)^{1/2}\ .
\label{RVM_norm_const}
\ee
The associated energy spectrum is given then by
\begin{equation}
\epsilon_n=n^{2} -\frac{b^{2}}{n^{2}}\, .
\label{RVM_spctr}
\end{equation}
Correspondingly, the wave functions for this case simplify and
are found as
\be
R_1(z)&=& \e^{-b z}\sin z\ C_1^{(0,b)}(\cot z)\ ,\nonumber\\
R_2(z)&=& \e^{-b z/2}\sin^2 z\ C_2^{(0,b)}(\cot z)\ ,\nonumber\\
\dots\nonumber\\
R_n(z)&=& \e^{-b z/n}\sin^n z\ C_n^{(0,b)}(\cot z)\ .
\label{psi6}
\ee
Wave functions for the first two (unnormalized) levels with
$a\not= 0$ are displayed in Figs.~III.
{\vspace*{1truecm}
\begin{figure}[htb]
\vskip 5.0cm
\includegraphics{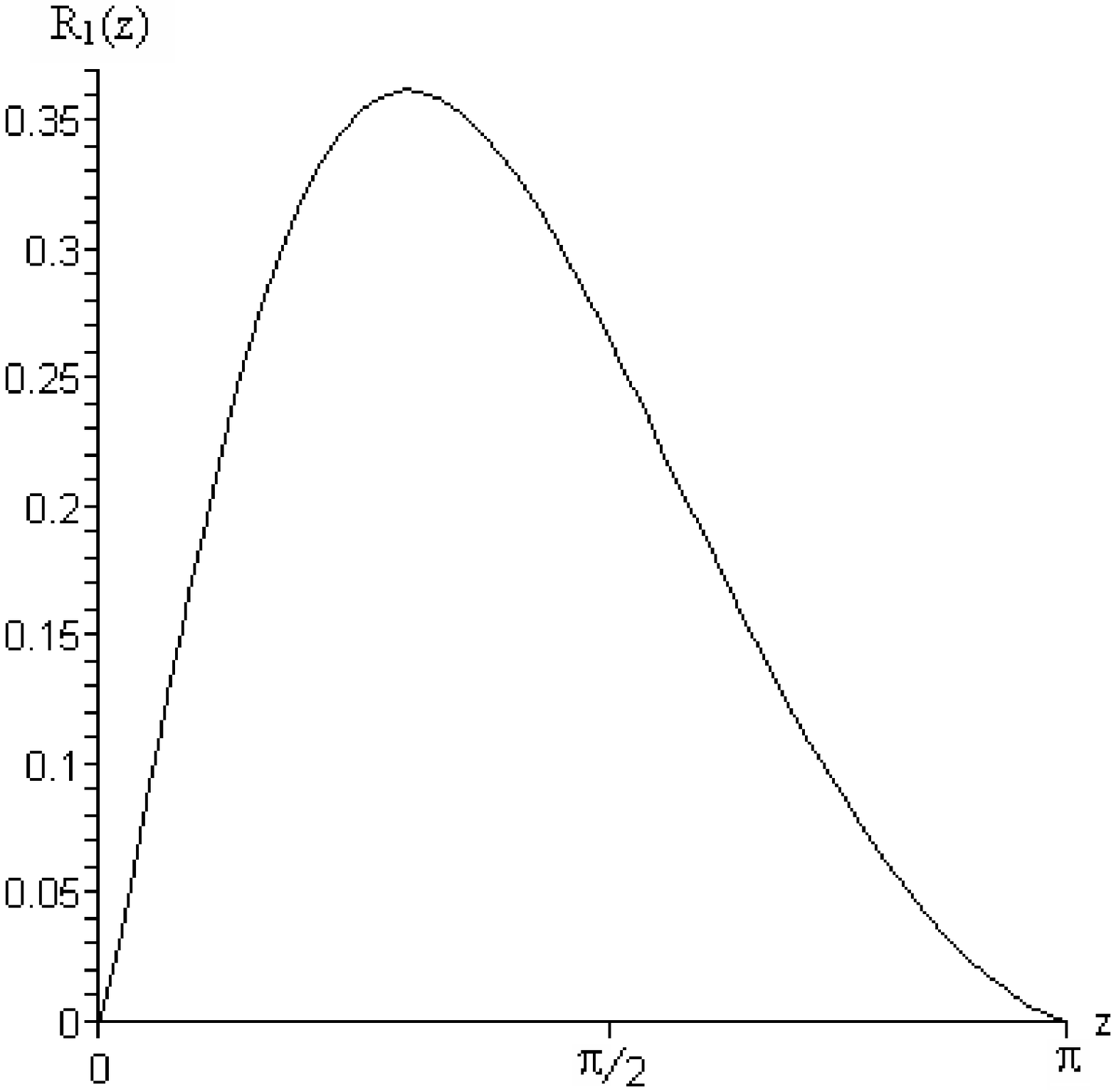}
\includegraphics{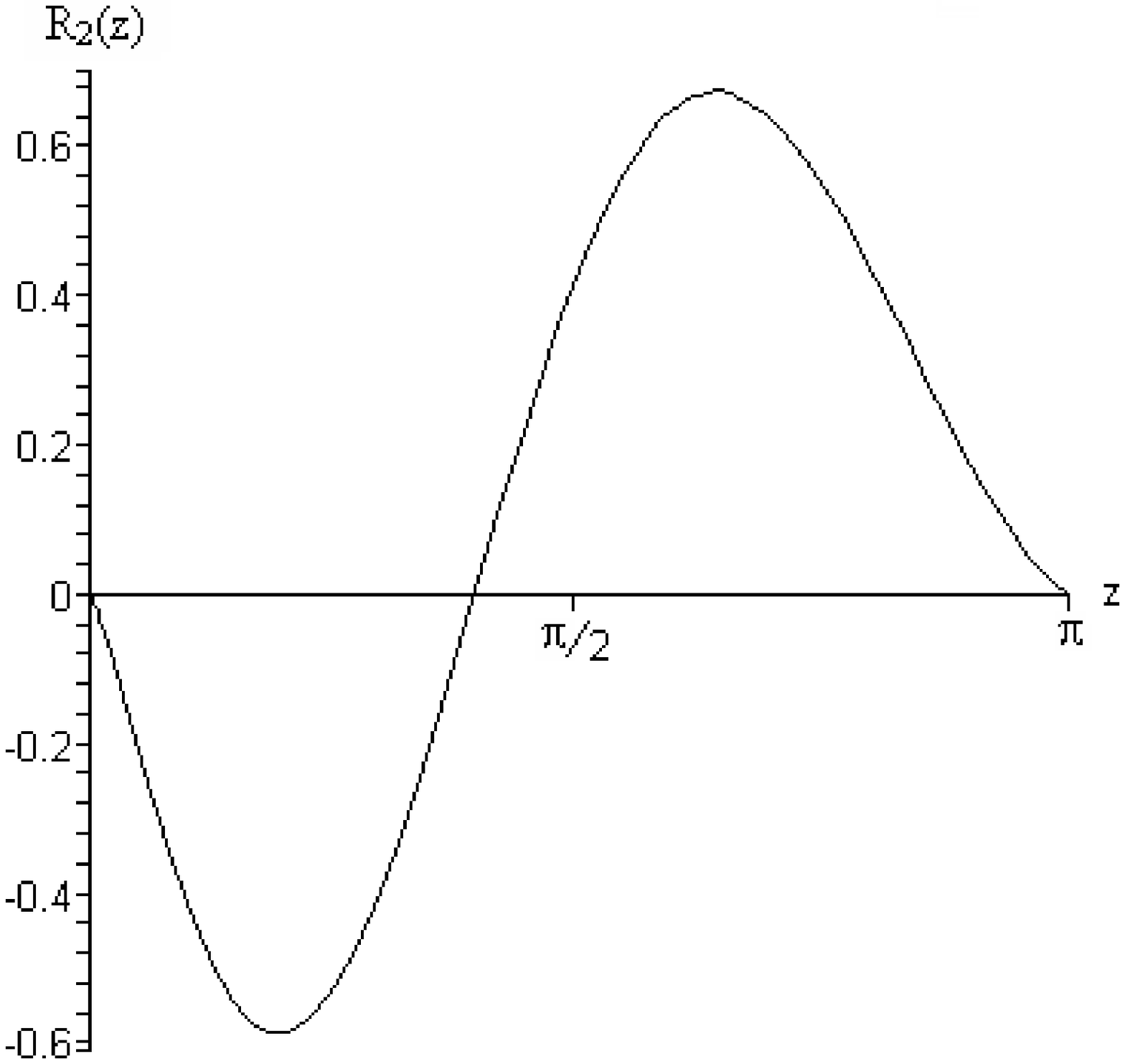}
\vspace{1.01cm}
{\small Fig.~III. 
Wave functions for the first two levels in the trigonometric
Rosen-Morse potential  for $a=$0.25 and $b=$1.
}\label{fig-psi}
\end{figure}


\section{The trigonometric Rosen-Morse superpotential.\label{C-susy-RMt}}
In this Section we derive the trigonometric Rosen-Morse superpotential
from the exact ground state solution which for $a\not=0$ has been
calculated as
\be R_1(z)\propto \e^{-bz/(a+1)}\sin^{a+1} z\ .\label{susy-R1-RMt} \ee
Substitution of the latter equation  (\ref{susy-R1-RMt}) 
into Eq.~(\ref{super_pt}) amounts to the following  superpotential,
\be U(z)= -{b \over a+1}+(a+1)\cot z\ ,\ee
shown in Fig~\ref{superpotencial}.
\begin{figure}[htbp]
\centerline{\psfig{figure=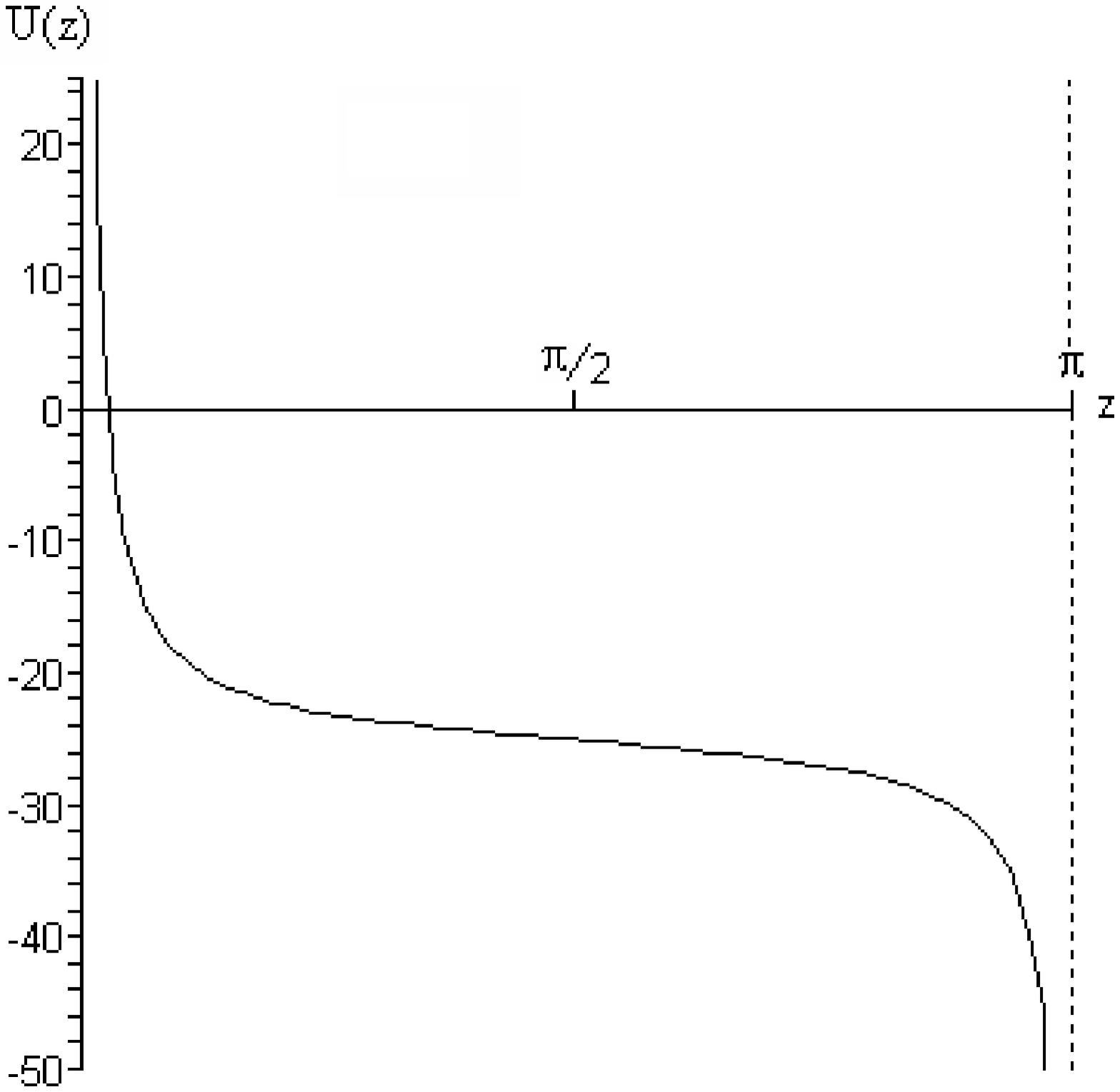,width=7cm}}
\vspace{0.1cm}
{\small Fig.~IV.
\hspace{0.2cm} The trigonometric Rosen-Morse superpotential.
The values of the displayed superpotential parameters 
are same as in Fig.~\ref{fig-RMt}.
}
\label{superpotencial}
\end{figure}
Correspondingly, the $A^\pm (z) $ operators are obtained as
\be A^\pm (z) = \pm {d\ \over dz}+(a+1)\cot z - {b \over a+1}\, .\ee
The corresponding Hamiltonian is then obtained identical to
Eq.~(\ref{v-RMt}), as should be.
The supersymmetric companion of $H(z)$, which is $\widetilde{H}(z)$,
becomes
\be \widetilde{H}(z) = -{d^{\, 2}\ \over dz^2}-2b\cot z +
(a+1)(a+2)\csc^2 z\, , \ee
and one finds the supersymmetric companions to the solutions as 
\be \widetilde R_n(z)|_{(a,b)}= R_{n-1}(z)|_{(a+1,b)}\ ,\ee
where  $n>1$, and  $R_{n-1}(z)|_{(a+1, b)}$
are among the exact solutions of the trigonometric 
Rosen-Morse potential.\\
As a further possible application of the solutions found here we wish
to mention the construction of  the so called
hierarchy of Hamiltonians ~\cite{Sukumar}, where 
one needs to have at ones disposal
exact orthonormalized functions  for all levels 
because in this case one  can pick up  any energy level,
$\epsilon_n$,  and its wave function $R_n(z)$. 
Obviously,  Eq.~(\ref{psi6}) fully qualify for that purpose.

\section{Concluding remarks.}
To recapitulate, the knowledge on the real exact solutions of the
one-dimensional Schr\"odinger equation for the trigonometric Rosen-Morse 
potential obtained in this work allows for a straightforward
construction of all the necessary ingredients of the supersymmetric 
quantum mechanics.
Compared to the text-book complex solutions, 
the exact real solutions reported here have the advantage to 
significantly simplify the calculations.

In addition, the trigonometric Rosen-Morse potential is more but just
one of the few exactly soluble quantum mechanical potentials. It joins 
as a new member the  smaller and important group of  potentials which 
generate physically relevant spectra.  
Indeed, in particle physics one encounters \cite{PL-B} 
the measured  excitation spectra of the nucleon $(N)$, 
and the so called $\Delta $ 
particle, in turn displayed in Fig.~\ref{fig-medido-ND}.

\begin{figure}[htb]
\vskip 5.0cm
\includegraphics{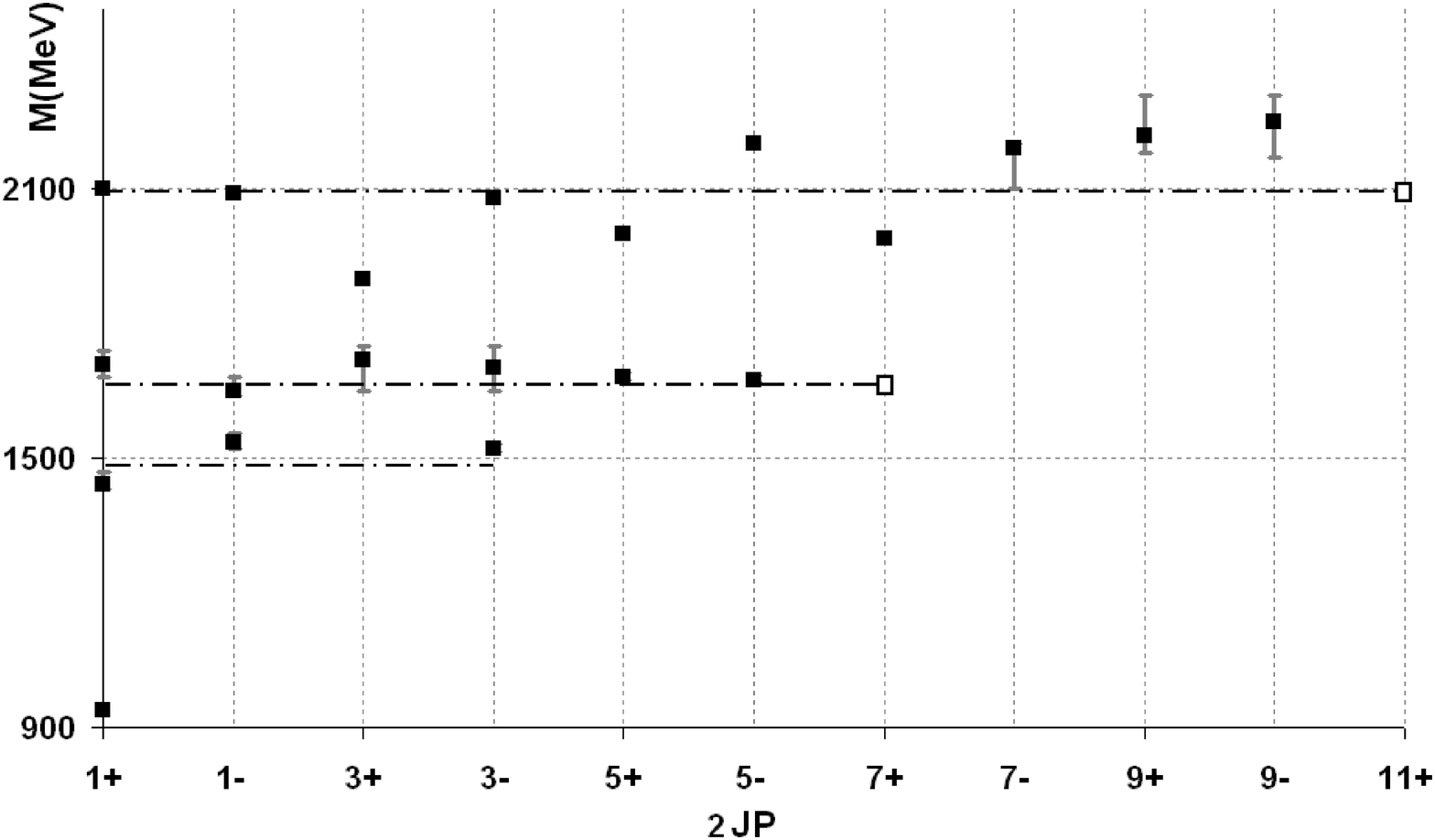}
\includegraphics{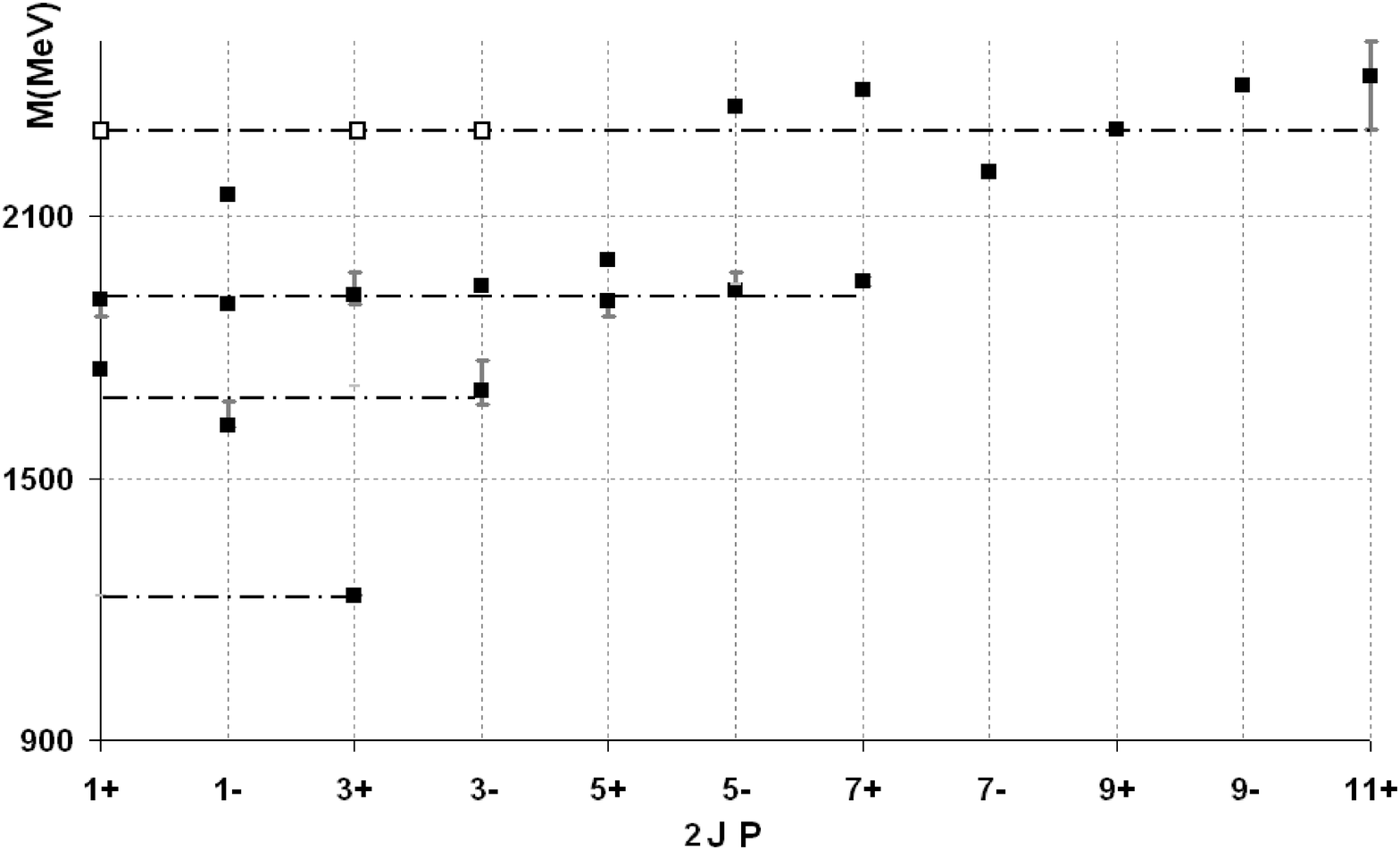}
\vspace{1.01cm}
{\small Fig.~V.
Experimentally observed baryon resonances (l.h.s.) $N$ and (r.h.s.) 
$\Delta$. The dash-point lines represent the mass average. 
Notice that the resonances
with masses above 2000 MeV are of significantly lower 
confidence but those with
masses below 2000 MeV where the degeneracy is very well pronounced. Empty 
squares denote predicted (``missing'') states.
}\label{fig-medido-ND}
\end{figure}

According to an observation due to Refs.~\cite{RVM-KMS}, \cite{RVM-K}
and references therein, those spectra repeat  with an amazing accuracy
the degeneracy patterns of the levels of the electron with spin 
in the hydrogen atom but are characterized by very
different mass splittings.
The following mass formula describes pretty well the averaged positions of
the three  narrow mass bands containing the
series of $(n-1)$ parity doubled states
(with $n=2,4$, and $6$) and  with spins ranging from $\frac{1}{2}^\pm $ to 
$\left( n-\frac{3}{2}\right)^\pm$ together with
the one unpaired  maximal spin $\left( n-\frac{1}{2}\right)^{P}$ state of 
either natural, or, unnatural parity, 
\begin{equation}
M_{(n;I)}-M^0_{(n;I)} =  g_{I}\frac{n^{2}-1}{4}-f_{I}\frac{1}{n^{2}}\, ,
\quad I=N, \Delta\, .
\label{mass-fla}
\end{equation}
Comparison of the baryon mass formula with Eq.~(\ref{RVM_spctr}) reveals  
coincidence between  the baryonic and the tRMP spectra.
The underlying constituent dynamics of the excited baryons
has been found to be  that of a quark--di-quark system \cite{RVM-KMS}.
In fact, in order to describe the particle spectra one needs to solve the
three dimensional Schr\"odinger equation but this does not cause much
 problems because  upon separation of the variables in polar coordinates
 the radial part of the three dimensional Schr\"odinger equation 
reduces to that very same one-dimensional Schr\"odinger equation 
(\ref{Sch-RMt}) up to the centrifugal barrier term, $l(l+1)/z^{2}$
(work in progress).
All in all, the trigonometric Rosen-Morse potential and its real 
orthogonal polynomial solutions open new venues in the calculation of 
interesting observables in
both supersymmetric quantum mechanics and particle spectroscopy. 

\section{Acknowledgments.}
We thank Dr.\ Alvaro P\'erez Raposo for his spontaneous and vivid
interest in our results. 
Work supported by Consejo Nacional de Ciencia y 
Technolog\'ia (CONACyT) Mexico under grant number C01-39280.

\pagebreak

\end{document}